\documentclass[twocolumn,showpacs,preprintnumbers,amsmath,amssymb]{revtex4}

\usepackage{graphicx}
\usepackage{dcolumn}
\usepackage{bm}

\begin{document}

\preprint{}

\title{Definitive observation of the dark triplet ground state \\ of charged excitons in high magnetic fields}

\author{G.~V.~Astakhov$^{1,2}$}
\author{D.~R.~Yakovlev$^{2,3}$}
\author{V.~V.~Rudenkov$^{4}$}
\author{P.~C.~M.~Christianen$^{4}$}
\author{T.~Barrick$^{5}$}
\author{S.~A.~Crooker$^{5}$}
\author{A.~B.~Dzyubenko$^{6}$}
\author{W.~Ossau$^{1}$}
\author{J.~C.~Maan$^{4}$}
\author{G.~Karczewski$^{7}$}
\author{T.~Wojtowicz$^{7}$}

\affiliation{ $^{1}$Physikalisches Institut der Universit\"{a}t
W\"{u}rzburg, 97074 W\"{u}rzburg, Germany \\
$^{2}$A.F.Ioffe Physico-Technical Institute, Russian
Academy of Sciences, 194021, St.Petersburg, Russia \\
$^{3}$Experimentelle Physik 2, Universit\"{a}t Dortmund, 44221
Dortmund, Germany \\
$^{4}$High Field Magnet Laboratory, University of Nijmegen, 6525 ED
Nijmegen, The Netherlands \\
$^{5}$National High Magnetic Field Laboratory, Los Alamos, New
Mexico 87545, USA \\
$^{6}$Department of Physics, California State University at
Bakersfield, Bakersfield, CA 93311, USA \\
$^{7}$Institute of Physics, Polish Academy of Sciences, PL-02668
Warsaw, Poland }

\date{\today}

\begin{abstract}
The ground state of negatively charged excitons (trions) in high
magnetic fields is shown to be a dark triplet state, confirming
long-standing theoretical predictions. Photoluminescence (PL),
reflection, and PL excitation spectroscopy of CdTe quantum wells
reveal that the dark triplet trion has lower energy than the singlet
trion above 24 Tesla. The singlet-triplet crossover is ``hidden"
({\em i.e.}, the spectral lines themselves do {\em not} cross due to
different Zeeman energies), but is confirmed by
temperature-dependent PL above and below 24\,T. The data also show
two bright triplet states.
\end{abstract}


\maketitle

A central problem found in atomic, solid state, and nuclear physics is the case of a three-particle
system of fermions, bound together by long-range Coulomb interactions.  In atomic physics, this
situation is most simply realized by the two-electron hydrogen ion, $H^{-}$, in which the two
identical electrons can exist in either a singlet or triplet state with total electron spin $S_e=0$
or $1$, depending on external parameters. The semiconductor analog of the $H^{-}$ ion is the
negatively charged exciton (trion), consisting of two conduction electrons bound to a single
valence hole. Optical signatures from trions have been observed in GaAs, CdTe, and ZnSe quantum
wells (QWs) \cite{Khe93,Fin95,Shi95,Ast99}. Unlike the $H^{-}$ ion, the hole and two electrons
comprising the trion have comparable masses and typically experience strong QW confinement in one
dimension, making trions a genuine quantum three-particle system with Coulomb interactions for
which no general analytical solutions exist.

%
%
Much attention has focused on the evolution of trion optical
signatures with applied magnetic field
\cite{Yus01,Vanhoucke02,Sch02,Nic02,Ash04}. In the limit of zero
magnetic field, theory predicts just one bound trion state: the
$S_e=0$ singlet trion ($T_{s}$) \cite{Ste89, Whi97}. This is
consistent with Hill's theorem \cite{Hill77}, which states that
the $H^-$ ion (with an infinitely massive proton) supports exactly
one bound singlet state. In the opposite limit of extremely high
magnetic fields, it can be rigorously shown that a $S_e=1$ triplet
is the only bound trion state in a strictly 2D system
\cite{Pal96,Whi97}. Model-independent symmetry considerations
\cite{Dzy00} demonstrate that this lowest triplet state is
``dark'' ($T_{td}$) ({\em i.e.}, optically inactive), due to the
exact selection rules imposed by spatial axial and translational
symmetries that exist in a disorder-free QW. Thus, at finite
magnetic fields one expects both singlet and triplet bound trions
\cite{Pal96, Whi97}. More importantly, at some critical magnetic
field $B_c$ the spin configuration of the trion ground state must
cross over from the singlet to the triplet. Theoretical estimates
suggest this crossover field is very large ($B_c>20$~T) and
depends sensitively on the strength of the Coulomb interaction
(dielectric constant) and the details of the QW confinement
\cite{Whi97, Woj00, Riv01,Yak01,Red02}. Numerical calculations
also point to the existence of weakly-bound, optically active
``bright" triplet states ($T_{tb}$), although there is large
disparity amongst the predicted regions of stability and binding
energies \cite{Woj00, Riv01}. Distinction between $T_{td}$ and
$T_{tb}$ is due to \textit{orbital motion} and \textit{is not
related} to the spin selection rules \cite{Dzy00,Woj00,Riv01}.
Note, $T_{td}$ and $T_{tb}$ have identical spin configuration.

\begin{figure*}[tbp]
\includegraphics[width=.81\textwidth]{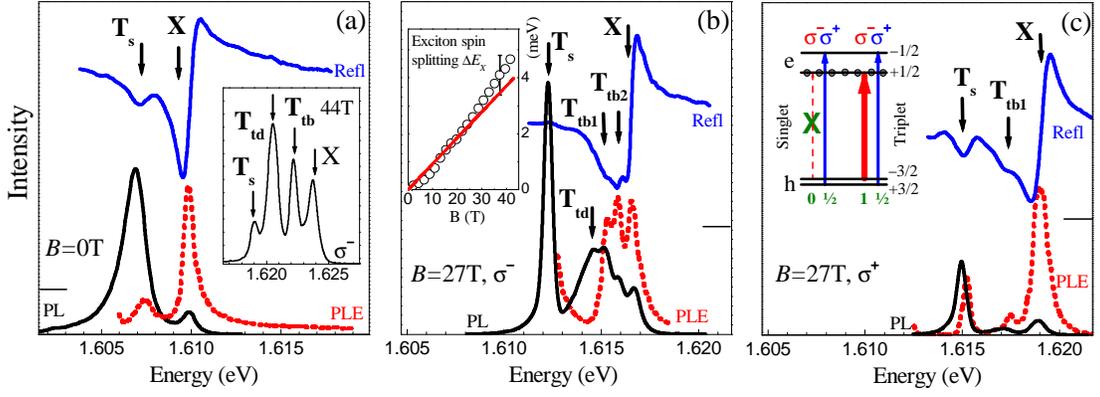}
\caption{ PL, PLE, and reflectivity (Refl) spectra of a 120\,{\AA}
CdTe/Cd$_{0.85}$Mg$_{0.15}$Te QW with $n_e =2 \times
10^{10}$\,cm$^{-2}$ at $T=1.3$\,K. The singlet ($T_{s}$), dark
triplet ($T_{td}$), two bright triplet ($T_{tb1}$ and $T_{tb2}$)
trions, and the neutral exciton ($X$) are clearly seen. (a)
Spectra at 0\,T. Inset: $\sigma ^{-}$ PL at $B=44$\,T. (b,c)
Spectra at 27\,T in $\sigma ^{-}$ and $\sigma ^{+}$ polarizations.
Inset (b): $X$ Zeeman splitting (circles) and electron splitting
(line) for $g_e=-1.60$. Inset (c): Schematic of transitions
leading to photocreation of $T_{s}$ and $T_{tb}$. Transition
probabilities are coded by the arrow thickness, dashed line means
the forbidden transition.} \label{fig12}
\end{figure*}

\begin{figure}[tbp]
\includegraphics[width=.21\textwidth]{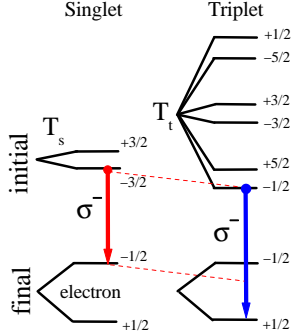}
\caption{ Schematic of the hidden singlet-triplet crossover of trion states in CdTe QWs. At high
magnetic fields, the lowest dark triplet state has less energy than the lowest singlet. After
recombination, however, the remaining electron is left in different spin states. Thus, the observed
triplet PL energy remains {\em larger\/} than that of the singlet, as indicated by the lengths of
the arrows.} \label{fig4}
\end{figure}

In this Letter we present conclusive evidence that the high-field
($B > B_c=24$\,T) ground state of negatively-charged trions in
CdTe-based QWs is 1) a triplet state, and 2) optically dark, --
{\em i.e.}, it has no absorption oscillator strength. Three
distinct and complementary polarization-resolved spectroscopies --
photoluminescence (PL), reflection, and PL excitation (PLE)
\cite{comment2} -- proved to be essential for identifying and
conclusively determining the spin properties of trions in magnetic
fields below {\em and above} the singlet-triplet crossover field
$B_c$. As such, this work represents what is to our knowledge the
first comprehensive picture of the evolution of the trion ground
state's spin over a complete range of magnetic fields. Two
important aspects of the singlet-triplet crossover, revealed
particularly in high-field PL spectra, require careful accounting
of the Zeeman energies of the initial trion and the final electron
states. First, the actual crossover point is shifted to much lower
fields ($B_c=24$\,T) than the $\sim$70\,T that is expected when
Zeeman energies are disregarded. Second, and less obvious, the
singlet-triplet crossing is {\em hidden\/} from direct observation
-- {\em i.e.}, the measured $T_{s}$ and $T_{td}$ PL peaks
themselves do {\em not\/} cross. This is because, following
emission, the spin (and therefore energy) of the final remaining
electron is different for $T_{s}$ and $T_{td}$. Rather, the
crossover is revealed by an exchange of {\em intensity\/} between
the $T_{s}$ and $T_{td}$ lines in PL, and by their temperature
dependence above and below $B_c$.

\begin{figure}[tbp]
\includegraphics[width=.32\textwidth]{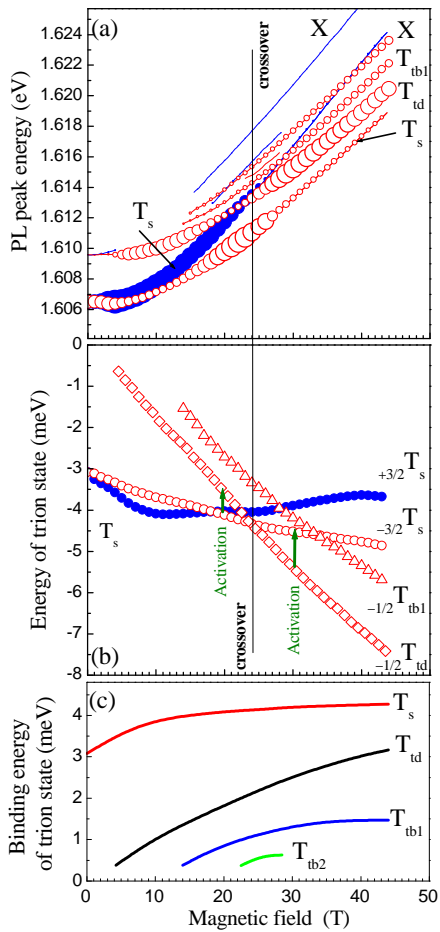}
\caption{(a) Evolution of trion PL peaks in $\sigma ^{-}$ (open)
and $\sigma ^{+}$ (solid) polarizations.  Symbol size indicates PL
intensity. (b) Energies of the $T_{s}$ (circles), $T_{td}$
(diamonds), and $T_{tb1}$ (triangles) trion states, measured from
the neutral exciton $X$, showing $B_c = 24$\,T. Arrows mark the
thermal activation processes measured in Fig.~\ref{temperature}.
(c) Trion binding energies from Coulomb interactions alone (no
Zeeman).} \label{fig3}
\end{figure}

Single 120\,{\AA} CdTe/Cd$_{0.85}$Mg$_{0.15}$Te QWs were grown by
molecular beam epitaxy on (100)-oriented GaAs substrates by
wedge-doping, which allows different electron densities $n_e$ on
the same wafer \cite{Greg}. The data presented here were from a
structure with $n_e=2\times 10^{10}$\,cm$^{-2}$. Polarized optical
spectra were measured at low temperatures (1.3--10\,K) and in
magnetic fields applied parallel to the growth axis (Faraday
geometry). DC fields to 33\,T (Nijmegen) and pulsed fields to
44\,T (Los Alamos) were used. Ti:Sapphire or He-Ne lasers with
power density $<$1\,W/cm$^{2}$ were used for excitation via fibers
or by direct optical access. Circularly polarized light was used
to resolve the exciton and trion spin orientation.

Zero-field optical spectra are shown in Fig.~\ref{fig12}a, where
the well-known pair of resonances associated with the neutral
exciton $X$ and singlet trion $T_s$ are clearly seen in PL, PLE,
and reflectivity. Triplet states, being unbound at zero field, are
not observed. Linewidths are $<$1.5\,meV, much smaller than the
3\,meV trion binding energy (taken as the energy difference
between $X$ and $T_s$ lines). The $T_s:X$ ratio of oscillator
strengths is 1:9 (from PLE and reflectivity), permitting
evaluation of the 2DEG density \cite{Ast02m}. At 44 Tesla (inset),
the PL spectra develop two additional strong peaks between the $X$
and $T_s$ lines, which we assign to dark and bright triplets based
on their polarization, energy, and evolution with magnetic fields
as discussed below.

In finite magnetic fields, correct assignment of the various
optical transitions to the proper exciton or trion state is
essential. Figures 1(b,c) show the polarized optical spectra at
27\,T. The neutral exciton $X$ is readily identified in
reflectivity, where it dominates all other resonances, exhibits
equal oscillator strength in both $\sigma ^{+}$ and $\sigma ^{-}$
polarizations, and also appears in an undoped reference sample.
The features observed at the same energy in PL and PLE spectra are
therefore also assigned to $X$. Note that while $X$ is strong in
PLE spectra, it is weak in PL due to thermalization to lower-lying
trion states.

At 27\,T, the $X$ Zeeman splitting is $\sim$2.7\,meV. The field
dependence of the $X$ and electron Zeeman splittings are shown in
the inset. The latter, determined by spin-flip Raman scattering
\cite{Sir97}, indicates an electron g-factor $g_e$=$-1.60$. The
exciton spin splitting, $\Delta E_X=(g_{hh} - g_e)\mu_B B$,
therefore implies a small heavy-hole g-factor ($|g_{hh}|<0.2$)
which actually changes sign at $\sim$18\,T.

Trion formation involves a photocreated electron-hole pair and a
background electron from the 2DEG. In high magnetic fields, when
the 2DEG is totally spin polarized ($B>4$\,T in this sample),
singlet and triplet trion states can be identified by their
distinct polarizations in PLE, reflectivity, and PL spectra. The
singlet trion $T_s$ with the lowest Zeeman energy has net spin
projection $S_z={S_{ez}+S_{hz}=0-\frac{3}{2}= -\frac{3}{2}}$ (see
Fig.~\ref{fig4}). This singlet emits a $\sigma ^{-}$ photon upon
recombination, leaving the remaining electron in the {\em upper}
(${-\frac{1}{2}}$) Zeeman state. Formation of $T_s$, requiring a
photoexcited electron with spin antiparallel to the 2DEG
electrons, should therefore exhibit a strong PLE and reflectivity
resonance {\em only} in $\sigma ^{+}$ polarization, as observed.
Indeed, $T_s$ is quite strong in $\sigma ^{+}$ PLE and
reflectivity, and completely absent in the $\sigma ^{-}$
reflectivity (its absence in $\sigma ^{-}$ PLE is obscured by
scattered light, where we always detect $\sigma ^{-}$ emission).

In contrast, triplet trions are predominantly polarized opposite
to the singlet. The lowest energy triplet has net spin
$S_z={S_{ez}+S_{hz}=+1-\frac{3}{2}= -\frac{1}{2}}$. This triplet
also emits a $\sigma ^{-}$ photon upon recombination, but unlike
$T_s$, leaves the remaining electron in the {\em lower}
(${+\frac{1}{2}}$) Zeeman state. This distinction will prove
important shortly, when discussing the hidden singlet-triplet
crossover. Formation of triplet trions, requiring predominantly
spin-parallel electrons, should therefore exhibit resonances
largely in the $\sigma ^{-}$ PLE and reflectivity.  Indeed, two
additional resonances in $\sigma ^{-}$ PLE and reflectivity are
clear, and both have corresponding $\sigma ^{-}$ PL emission
(Fig.~\ref{fig12}b). We therefore assign these lines to two
``bright'' (optically active) triplet trion states $T_{tb1}$ and
$T_{tb2}$. Most importantly, however, an additional strong $\sigma
^{-}$ polarized PL peak is seen at energy 1.6145\,eV. It has {\em
no counterpart\/} in PLE or reflectivity spectra, meaning that the
corresponding transition has no oscillator strength and is
optically inactive. Thus, we assign this PL peak to the dark
triplet $T_{td}$. It has the largest binding energy among the
triplet states, consistent with theoretical predictions
\cite{Whi97,Woj00,Riv01}. The reason dark triplet PL appears at
all is due to the small but nonzero probability of allowed
radiative recombination via disorder scattering \cite{Dzy00} or
interaction with excess electrons, as demonstrated even in low
density 2DEGs ($n_e \sim 10^{10}$\,cm$^{-2}$) \cite{San02}.

Figure 3a shows the energy shifts of the trion PL with magnetic
field, where symbol size indicates the PL intensity and weak
transitions are traced by lines. We concentrate primarily on the
evolution of the $T_s$ and $T_{td}$ peaks.  For all accessible
fields (0--44\,T), the $T_s$ PL peak occurs at the lowest measured
energy.  However, at about 24\,T, the $T_s$ PL intensity is
significantly redistributed in favor of $T_{td}$, strongly
suggesting that the bound dark triplet has crossed the singlet to
become the trion ground state. However, the observed PL lines
themselves {\em do not cross}. This seeming contradiction is
resolved by recalling that the electrons which remain after
recombination of $T_s$ and $T_{td}$ reside in the upper
(${-\frac{1}{2}}$) and lower (${+\frac{1}{2}}$) spin states
respectively, and these final states are split by the electron
Zeeman energy $\Delta E_e = \mu _B |g_e| B$.  As shown
schematically in Fig. 2, the $T_{td}$ state can have lower energy
than $T_s$, but emission from $T_{td}$ may still have the greater
energy. At the crossover field $B_c$, when the $T_s$ and $T_{td}$
states themselves have identical energies, the energy of $T_{td}$
emission still exceeds the energy of $T_s$ emission by exactly
$\Delta E_e$.  We describe the change of trion ground state as a
``hidden" crossing between $T_s$ and $T_{td}$ \cite{comment3}.

\begin{figure}[tbp]
\includegraphics[width=.36\textwidth]{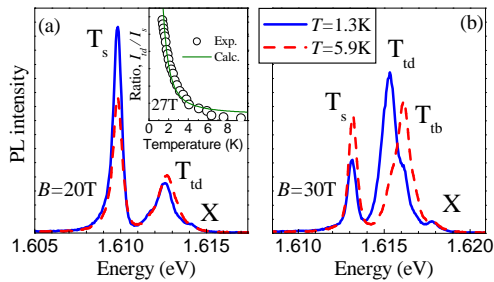}
\caption{The temperature dependence of PL spectra measured in
$\sigma ^{-}$ polarization at 20\,T (a) and 30\,T (b). Inset: the
$T_{td}$:$T_{s}$ intensity ratio vs. temperature. The line is a
calculation based on a two-level model with 0.45\,meV energy
splitting.} \label{temperature}
\end{figure}

The hidden crossover is revealed particularly well by the
temperature dependence of the trion PL peaks above and below
$B_c=24$\,T (Fig.~\ref{temperature}).  At 20\,T (below $B_c$),
increasing the temperature from 1.3 to 5.9\,K depopulates the
$T_s$ state in favor of $T_{td}$, implying thermal excitation of
trions from a singlet ground state to a higher-energy dark
triplet. In contrast, the same temperature increase at 30\,T
(above $B_c$) has the opposite effect -- an increase in $T_s$
emission and a reduction in $T_{td}$ emission, implying thermal
excitation from a dark triplet ground state to a higher-lying
singlet state.  In other words, the trion ground state has crossed
over from singlet to dark triplet.  A fit to the ratio of PL
intensities vs. temperature (the inset of Fig.~\ref{temperature})
reveals that the radiative recombination times of the trion states
satisfy $t_{td} \gg t_{s}$ \cite{value}, independently confirming
the identification of $T_{td}$ as a dark state.

Whereas the Zeeman splitting of the final electron states causes
the ``hidden" nature of the crossover, the different Zeeman
splittings of the initial $T_s$ and $T_{td}$ states has an
additional important consequence.  Namely, the crossover occurs at
a much lower magnetic field than it would in the absence of Zeeman
effects. Figure 3b shows the {\em initial}  energies of all trion
states, measured with respect to the ``center-of-gravity'' of the
neutral exciton Zeeman doublet, which accounts for the overall
diamagnetic shift. Each trace has a contribution from the trion's
Coulomb binding energy, as well as the additional Zeeman energy of
the initial trion state. The striking feature of Fig.~\ref{fig3}b
is the evident crossover between $T_{td}$ (with spin projection
$S_z=-\frac{1}{2}$) and $T_{s}$ (with $S_z = -\frac{3}{2}$) which
occurs at 24\,T. This value coincides very well with the field at
which the PL intensity redistribution occurs. Note also that a
bright triplet state $T_{tb1}$ (with $S_z = -\frac{1}{2}$) crosses
the singlet at $\sim$34\,T. For future comparison with theory, we
also plot in Fig.~\ref{fig3}c the trion binding energies resulting
from Coulomb interactions alone (i.e., {\em without} Zeeman
terms). It is evident that the actual crossover field is indeed
reduced due to the electron Zeeman splitting $\Delta E_e$, without
which $B_c$ would estimated to $\sim$70\,T, in good qualitative
agreement with theoretical predictions for II-VI QWs \cite{Yak01}.

As discussed briefly above, the data also reveal another novel
feature: a second bright triplet state $T_{tb2}$. It is detected
in PL, PLE, and reflectivity spectra between 22\,T and 28\,T (see
Figs.~\ref{fig12}b and \ref{fig3}c). No experimental observations
of this state have been reported to date. While CdTe QWs are
characterized by strong Coulomb interactions, this enhancement is
not enough to ensure binding of additional trion states because
the neutral exciton -- relative to which a trion may or may not be
bound -- is also more tightly bound. Some new physics is needed
here. One possibility is that trion binding energies are enhanced
in these QWs by ``bipolaron" effects, wherein the polarization
clouds of two electrons in the trion partly overlap, lowering the
total energy relative to the neutral exciton $X$ \cite{Riv00}.

In conclusion, combined PL, PLE, and reflectivity studies reveal
the detailed energy spectrum of charged trions over a wide range
of magnetic fields. These trions exemplify a canonical problem of
interest in many solid state, atomic, and nuclear physics
problems: a three-particle spin system with long-range Coulomb
interactions. We have confirmed a high-field crossover from the
singlet to dark triplet trion state, upholding long-standing
theoretical predictions. It has been shown that the Zeeman spin
splitting of electrons both reduces the crossover field to
experimentally accessible values $B_c=24$\,T, and also causes the
crossover to be ``hidden" from direct observation of the emission
energies themselves. We have also observed a novel feature in the
spectra, an additional bound bright triplet trion state, and
indicated the physics that might explain its stability.

This work was supported  by the the Deutche Forschungsgemeinschaft
(SFB 410), the DARPA Spins program, the NSF grant DMR-0203560, and
the EU IHP-ARI program HPRI-CT-1999-00036.



\begin{references}

\bibitem{Khe93} K.~Kheng \textit{et al.}, Phys. Rev. Lett. \textbf{71}, 1752 (1993).

\bibitem{Fin95} G.~Finkelstein \textit{et al.},
Phys. Rev. Lett. \textbf{74}, 976 (1995).

\bibitem{Shi95} A.J.~Shields \textit{et al.}, Phys. Rev. B \textbf{52}, 5523(R) (1995).

\bibitem{Ast99} G.V.~Astakhov \textit{et al.}, Phys. Rev. B
\textbf{60}, 8485(R) (1999).

\bibitem{Yus01} G.~Yusa  \textit{et al.},
Phys. Rev. Lett. \textbf{87}, 216402 (2001).

\bibitem{Vanhoucke02} T. Vanhoucke \textit{et al.}, Phys. Rev.
B \textbf{65}, 233305 (2002).

\bibitem{Sch02} C.~Sch\"{u}ller  \textit{et al.},
Phys. Rev. B \textbf{65}, 081301(R) (2002).

\bibitem{Nic02}H.~A.~Nickel  \textit{et al.}
                  Phys. Rev. Lett. {\bf 88}, 056801 (2002).

\bibitem{Ash04} B.M.~Ashkinadze \textit{et al.}, Phys. Rev. B \textbf{69}, 115303
(2004).

\bibitem{Ste89} B.~Stebe  \textit{et al.}, Superlatt. Microstruct.
\textbf{5}, 545 (1989).

\bibitem{Whi97} D.M.~Whittaker  \textit{et. al.}, Phys. Rev. B \textbf{56}, 15185 (1997).

\bibitem{Hill77}R.~N.~Hill, Phys. Rev. Lett. {\bf 38}, 643 (1977).

\bibitem{Pal96} J.J.~Palacios  \textit{et. al.},
Phys. Rev. B \textbf{54}, 2296(R) (1996).

\bibitem{Dzy00} A.B.~Dzyubenko  \textit{et al.}, Phys. Rev. Lett. \textbf{84}, 4429 (2000);
Physica E \textbf{6}, 226 (2000).

\bibitem{Woj00} A.~Wojs  \textit{et al.}, Phys. Rev. B \textbf{62}, 4630 (2000);
Physica E \textbf{8}, 254 (2000).

\bibitem{Riv01} C.~Riva  \textit{et al.}, Phys. Rev. B \textbf{63}, 115302
(2001).

\bibitem{Yak01} D.R.~Yakovlev \textit{et al.}, Phys. Stat. Sol. (b) \textbf{227}, 353
(2001).

\bibitem{Red02} P.~Redli\'{n}ski \textit{et al.}, Semicond. Sci. Technol. \textbf{17},
237 (2002).

\bibitem{comment2} To our knowledge, the absorption spectra of
the trion triplet was only studied in Ref.~\onlinecite{Sch02}. However, the spectra were taken
using unpolarized light leading to the loss of all the relevant information concerning the Zeeman
splittings. This might strongly affect the interpretation of experimental data.

\bibitem{Greg} T.~Wojtowicz \textit{et al.}, Acta
Physica Polonica A \textbf{94}, 199 (1998).

\bibitem{Ast02m} G.V.~Astakhov \textit{et al.}, Phys. Rev. B \textbf{65}, 115310
(2002).

\bibitem{Sir97} A.A.~Sirenko \textit{et al.}, Phys. Rev. B \textbf{56}, 2114 (1997).

\bibitem{comment3} Spin structure of trions in magnetic fields (see e.g.
Fig.~\ref{fig4}) determine whether the singlet-triplet crossing is
hidden or not hidden. It depends on the ratio of absolute values
and signs of electron and hole g-factors, which vary with material
and structure parameters.

\bibitem{San02} D.~Sanvitto \textit{et al.}, Phys. Rev. Lett. \textbf{89}, 246805
(2002).

\bibitem{value} The exact values used to fit the experimental data
in the inset of Fig.~\ref{temperature} are $t_{td}=200t_{s}$ and
$\tau = 4t_{s}$, where $\tau$ is the relaxation time between the
$T_{td}$ and $T_{s}$ states.


\bibitem{Riv00} C.~Riva \textit{et al.}, Phys. Rev. B \textbf{61}, 13873 (2000).

\end{references}
\end{document}